\documentclass[%
 reprint,
 superscriptaddress,
 bibnotes,
 amsmath,amssymb,
 aps,
 prb,
 floatfix,
]{revtex4-2}
\usepackage{graphicx}
\usepackage{mathptmx,bm}
\begin{document}

%\preprint{APS/123-QED}

\title{Magnetic Properties and Electronic Configurations of Mn Ions in the Diluted Magnetic Semiconductor Ba$_{1-x}$K$_{x}$(Zn$_{1-y}$Mn$_{y}$)$_{2}$As$_{2}$ Studied by X-ray Magnetic Circular Dichroism and Resonant Inelastic X-ray Scattering }
\author{H. Suzuki}

\affiliation{Department of Physics, University of Tokyo, Bunkyo-ku, Tokyo 113-0033, Japan}
\affiliation{Frontier Research Institute for Interdisciplinary Sciences, Tohoku University, Sendai, 980-8578, Japan}
\affiliation{Institute of Multidisciplinary Research for Advanced Materials (IMRAM), Tohoku University, Sendai 980-8577, Japan}
\author{G. Q. Zhao}
\affiliation{Beijing National Laboratory for Condensed Matter Physics, Institute of Physics, Chinese Academy of Sciences, Beijing 100190, China}
\author{J. Okamoto}
\affiliation{National Synchrotron Radiation Research Center, Hsinchu 30076, Taiwan}
\author{S. Sakamoto}
\affiliation{Department of Physics, University of Tokyo, Bunkyo-ku, Tokyo 113-0033, Japan}
\affiliation{Institute for Solid State Physics (ISSP), University of Tokyo, Kashiwa, Chiba 277-8581, Japan}

\author{Z.-Y. Chen}
\affiliation{National Synchrotron Radiation Research Center, Hsinchu 30076, Taiwan}

\author{Y. Nonaka}
\affiliation{Department of Physics, University of Tokyo, Bunkyo-ku, Tokyo 113-0033, Japan}

\author{G. Shibata}
\affiliation{Department of Physics, University of Tokyo, Bunkyo-ku, Tokyo 113-0033, Japan}
\affiliation{Department of Applied Physics, Tokyo University of Science, Katsushika-ku, Tokyo 125-8585, Japan}

\author{K. Zhao}
\affiliation{Beijing National Laboratory for Condensed Matter Physics, Institute of Physics, Chinese Academy of Sciences, Beijing 100190, China}

\author{B. J. Chen}
\affiliation{Beijing National Laboratory for Condensed Matter Physics, Institute of Physics, Chinese Academy of Sciences, Beijing 100190, China}

\author{W.-B. Wu}
\affiliation{National Synchrotron Radiation Research Center, Hsinchu 30076, Taiwan}

\author{F.-H. Chang}
\affiliation{National Synchrotron Radiation Research Center, Hsinchu 30076, Taiwan}

\author{H.-J. Lin}
\affiliation{National Synchrotron Radiation Research Center, Hsinchu 30076, Taiwan}

\author{C.-T. Chen}
\affiliation{National Synchrotron Radiation Research Center, Hsinchu 30076, Taiwan}

\author{A. Tanaka}
\affiliation{Department of Quantum Matter, ADSM, Hiroshima University, Higashi-Hiroshima 739-8530, Japan}

\author{M. Kobayashi}
\affiliation{Center for Spintronics Research Network, The University of Tokyo, 7-3-1 Hongo, Bunkyo-ku, Tokyo 113-8656, Japan}
\affiliation{Department of Electrical Engineering and Information Systems, The University of Tokyo, Bunkyo, Tokyo 113-8656, Japan}

\author{Bo Gu}
\affiliation{Advanced Science Research Center, Japan Atomic Energy Agency, Tokai 319-1195, Japan}
\affiliation{Kavli Institute for Theoretical Sciences, University of Chinese Academy of Sciences, Beijing 100190, China}

\author{S. Maekawa}
\affiliation{Advanced Science Research Center, Japan Atomic Energy Agency, Tokai 319-1195, Japan}
\affiliation{Kavli Institute for Theoretical Sciences, University of Chinese Academy of Sciences, Beijing 100190, China}
\affiliation{RIKEN Center for Emergent Matter Science (CEMS), Saitama 351-0198, Japan}

\author{Y. J. Uemura}
\affiliation{Department of Physics, Columbia University, New York, New York 10027, USA}

\author{C. Q. Jin}
\affiliation{Beijing National Laboratory for Condensed Matter Physics, Institute of Physics, Chinese Academy of Sciences, Beijing 100190, China}
\affiliation{Collaborative Innovation Center of Quantum Matter, Beijing, China}

\author{D. J. Huang}
\affiliation{National Synchrotron Radiation Research Center, Hsinchu 30076, Taiwan}

\author{A. Fujimori}
\affiliation{Department of Physics, University of Tokyo, Bunkyo-ku, Tokyo 113-0033, Japan}
\affiliation{National Synchrotron Radiation Research Center, Hsinchu 30076, Taiwan}
\affiliation{Department of Applied Physics, Waseda University, Shinjuku-ku, Tokyo 169-8555, Japan}

\date\today

\begin{abstract}
The magnetic properties and the electronic excitations of the new diluted magnetic semiconductor Ba$_{1-x}$K$_{x}$(Zn$_{1-y}$Mn$_{y}$)$_{2}$As$_{2}$ have been studied by x-ray magnetic circular dichroism (XMCD) and resonant inelastic x-ray scattering (RIXS) at the Mn $L_{2,3}$ edge. The sum rule analysis of the XMCD spectra yields the net spin moment of $0.45\mu_{\text{B}}$/Mn and the small orbital moment of $0.05\mu_{\text{B}}$/Mn. This indicates that the Mn atoms are in the high-spin configurations of $d^{5}$, whereas the presence of competing ferromagnetic and antiferromagnetic interactions between the Mn ions reduces the net spin moment. RIXS spectra show broad peaks from 1  to 6 eV energy loss, which originate from the $d$-$d$ crystal field excitations of the Mn ions.  Based on a  comparison of the RIXS line shapes with those of Ga$_{1-x}$Mn$_{x}$As, we conclude that the ground state of Mn in Ba$_{1-x}$K$_{x}$(Zn$_{1-y}$Mn$_{y}$)$_{2}$As$_{2}$ consists not only of the charge-transferred $3d^{5}\underline{L}$ electron configuration ($\underline{L}$: ligand hole) with weakly bound holes as in Ga$_{1-x}$Mn$_{x}$As, but also of the pure $3d^{5}$ configuration with free holes.

%pure acceptor $3d^{5}$ and charge-transferred $3d^{5}\underline{L}$ electron configurations constitute the ground state of Mn in Ba$_{1-x}$K$_{x}$(Zn$_{1-y}$Mn$_{y}$)$_{2}$As$_{2}$.%From configuration-interaction cluster-model calculations, we assign them to $d$-$d$ excitations from the mixture of Mn$^{2+}$ and Mn$^{3+}$ ground states. The broadness originates from the rapid decay of the $d$-$d$ excitons to electron-hole pairs created in the As $4p$ valence and conduction bands of the host material, which has small indirect band gap of 0.3 eV.
\end{abstract}

%\pacs{75.50.Pp,78.20.Ls,61.05.C-,78.70.Dm}%Pacs changed. Nov 19 2015
% Classification Scheme. %\keywords{Suggested keywords}%Use show keys class option if keyword

\maketitle

 %General introduction of DMS
Diluted magnetic semiconductors (DMSs) are promising candidates for future spintronic devices due to the possibility of utilizing both charge and spin degrees of freedom in a compatible way with the existing semiconductor technologies \cite{Ohno.H_etal.Science1998,Dietl.T_etal.Nat-Mater2010,Zutic.I_etal.Rev_mod_Phys2004,Ohno.H_etal.Applied-Physics-Letters1996,Dietl.T_etal.Science2000,Dietl.T_etal.Rev.-Mod.-Phys.2014}. Possible applications include nonvolatile semiconductor memories, spin-polarized light-emitting diodes \cite{Holub.M_etal.J.-Phys.-D:-Appl.-Phys.2007}, and tunnel magnetoresistance devices \cite{Tanaka.M_etal.Phys.-Rev.-Lett.2001}. In particular, the discovery of ferromagnetism in Mn-doped GaAs (GaMnAs) with the ferromagnetic Curie temperature ($T_{\text{C}}$) up to 185 K \cite{Novak.V_etal.Phys.-Rev.-Lett.2008,Wang.M_etal.Appl.-Phys.-Lett.2008} increased the possibility \cite{Jungwirth.T_etal.Rev.-Mod.-Phys.2014}. However, although various kinds of ferromagnetic semiconductors in thin-film form have been fabricated to achieve higher $T_{\text{C}}$'s and novel functionalities, the growth of specimens with full control of carrier density and magnetism has been a technological challenge. In GaMnAs, for example, the chemical solubility of the magnetic Mn$^{2+}$ ions remains low due to the valence mismatch with the Ga$^{3+}$ ions and the carrier density is difficult to control due to defect formation. Furthermore, unavoidable interstitial Mn ions couple antiferromagnetically with the substitutional Mn ions, deteriorating the long-range ferromagnetic order. From scientific points of view, it is still under debate whether the metallic transport is caused by the exchange-split host valence band as described by the Zener $p$-$d$ exchange model \cite{Dietl.T_etal.Science2000} or by an impurity band derived from nearly localized Mn $3d$ holes \cite{Chapler.B_etal.Phys.-Rev.-B2011}. Correlation between  $T_{\text{C}}$ and the presence of nanodecompositions under different growth conditions also needs to be clarified \cite{Dietl.T_etal.Rev.-Mod.-Phys.2015}.

The newly-synthesized DMS Ba$_{1-x}$K$_{x}$(Zn$_{1-y}$Mn$_{y}$)$_{2}$As$_{2}$ (Mn-BaZn$_{2}$As$_{2}$) has the same crystal structure as the "122"-type iron-based superconductors such as substituted BaFe$_{2}$As$_{2}$ and has a $T_{\text{C}}$ as high as 230 K \cite{Zhao.K_etal.Nat-Commun2013,Zhao.K_etal.Chin.-Sci.-Bull.2014}. The advantage of this material is that the charge-reservoir Ba layer and the magnetic ZnAs layer are separate, which allows the independent control of the amount of hole carriers and magnetic ions by substituting K for Ba and Mn for Zn, respectively. Additionally, the substitution of the divalent Mn ions for the isovalent Zn ions allows one to avoid the low chemical solubility and the phase segregation problems encountered in GaMnAs. Since single crystals are now available and the interstitial Mn incorporation is energetically precluded \cite{Glasbrenner.J_etal.Phys.-Rev.-B2014}, Mn-BaZn$_{2}$As$_{2}$ can also serve as a suitable system to study the magnetic impurity states and their interactions with carriers that mediate ferromagnetic correlations between the local spins. Our previous x-ray photoemission study on polycrystalline samples has shown that the Mn atoms take $S=5/2$ spin states and that the Mn $3d$ partial density of states is similar to that of GaMnAs \cite{Suzuki.H_etal.Phys.-Rev.-B2015}, reflecting the similar local coordination of the As atoms to the Mn atoms. On the other hand, we have also found, by soft-x-ray angle-resolved photoemission spectroscopy on single crystals, that the Mn $3d$ impurity band is formed near the Fermi level ($E_{F}$) and is located well below the valence band maximum (VBM) \cite{Suzuki.H_etal.Phys.-Rev.-B2015_ARPES} unlike in GaMnAs, where the impurity band is formed slightly above the VBM. As an effort toward device applications, Xiao \textit{et. al.} \cite{Xiao.Z_etal.Thin-Solid-Films2014} reported the growth of epitaxial thin films of the host semiconductor BaZn$_{2}$As$_{2}$ on MgO (001) substrates, which may enable the fabrication of BaZn$_{2}$As$_{2}$-based functional heterostructures. In order to realize spintronic devices, more detailed information about the electronic and magnetic properties of the Mn impurity states in Mn-BaZn$_{2}$As$_{2}$ is necessary.

In the present work, we have performed x-ray magnetic circular dichroism (XMCD) and resonant inelastic x-ray scattering (RIXS) measurements on Mn-BaZn$_{2}$As$_{2}$ ($x=0.3$, $y=0.15$, $T_{\text{C}}$ = 60 K) single crystals. XMCD directly probes the ferromagnetic component of magnetic moments in the ground states and RIXS allows us to investigate electronic excitations with element specificity. Here, we study the magnetic states and electronic configuration of Mn by utilizing x-ray photons around the Mn $L_{2,3}$ edge. 

Mn-BaZn$_{2}$As$_{2}$ single crystals were synthesized by the method reported in Ref. \onlinecite{Zhao.G_etal.Sci.-Rep.2017}. XMCD measurements were performed at the Dragon Beamline BL-11A of Taiwan Light Source at National Synchrotron Radiation Research Center (NSRRC), Taiwan. X-ray absorption spectroscopy (XAS) and XMCD spectra were taken at 30 K in the total-electron yield (TEY) mode with the probing depth of $\sim$ 5 nm. The XMCD spectrum was obtained by switching the direction of the external magnetic field ($\pm 1$ T). The monochromator resolution was $E/\Delta E>10000$. Single crystals oriented along the $c$-axis ($2$ mm $\times$ $2$ mm) were cleaved \textit{in situ} prior to the measurements to obtain fresh surfaces. External magnetic field was applied along the $c$ axis and the angle between the incident light and the magnetic field was 30$^{\circ}$. RIXS measurements were performed at the undulator beam line BL-05A1 of NSRRC. The optical layout of the present RIXS system with an active-grating monochromator and an active-grating spectrometer designed for the efficient collection of inelastically scattered x-rays is described in Ref. \onlinecite{Lai.C_etal.J.-Synchrotron-Radiat.2014}. The zero energy loss line was determined from the peak position of the elastic scattering peak of reference polycrystalline carbon and the total energy resolution around the Mn $L_{3}$ edge estimated from its full width half maximum was 80 meV ($E/\Delta E\simeq 8000$).  The RIXS spectra were measured with two linear polarizations at 20 K. % under a vacuum of 1.0 $\times$ 10$^{-9}$ Torr.
The polarizations of the scattered x-ray photons were not analyzed. 

%%XMCD%%
\begin{figure}[htbp] %  figure placement: here, top, bottom, or page
   \centering
   \includegraphics[width=8cm]{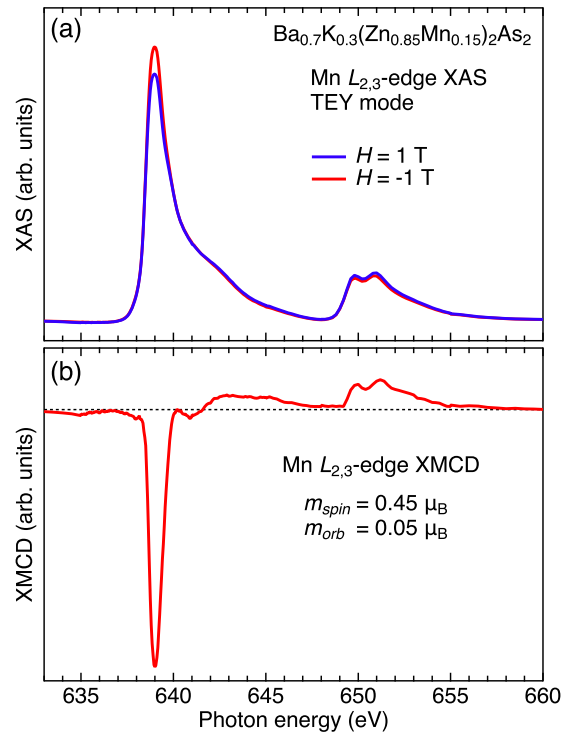} 
%   \caption{(Color online) (a) Valence-band ARPES difference spectrum of Mn-BaZn$_{2}$As$_{2}$. The off-resonance ARPES spectrum taken at 637 eV is subtracted from the on-resonance one at 640 eV in order to highlight the resonance enhancement of spectral weight. No significant enhancement due to the Mn $3d$ impurity band was observed near $E_{F}$. (b) Mn $3d$ partial density of states obtained by integrating panel (a) over momenta. Off- and on-resonance photoemission spectra obtained from Figs. \ref{RPES} (b) and (c) are also shown.}
   \caption{(Color online) (a) X-ray absorption spectra of Ba$_{0.7}$K$_{0.3}$(Zn$_{0.85}$Mn$_{0.15}$)$_{2}$As$_{2}$ single crystals at the Mn $L_{2,3}$ edge taken in the total-electron-yield mode. Magnetic fields of $\pm$ 1 T were applied along the $c$ axis, and x-rays were incident 30$^{\circ}$ from the $c$-axis. (b) X-ray magnetic circular dichroism at the Mn $L_{2,3}$ edge. The spin and orbital moments per Mn ion are deduced to be 0.45 and 0.05 $\mu_{\text{B}}$, respectively.}
   \label{XMCD}
\end{figure}

Figure \ref{XMCD} (a) shows the Mn $L_{2,3}$-edge XAS spectra of Mn-BaZn$_{2}$As$_{2}$. Magnetic fields of $\pm$ 1 T were applied along the $c$-axis. The single peak structure of the $L_{3}$ main peak evidences that Mn-BaZn$_{2}$As$_{2}$ does not contain extrinsic Mn components such as oxidized Mn atoms at the surface unlike in the case of GaMnAs \cite{Takeda.Y_etal.Phys.-Rev.-Lett.2008}. 
The XMCD spectrum is shown in Fig. \ref{XMCD} (b). The XMCD lineshape is consistent with a previous report for samples with slightly different doping levels \cite{Sakamoto.S_etal.ACS-Appl.-Electron.-Mater.2021}. From the experimental XMCD spectra, we have determined the spin and orbital magnetic moments separately by applying the XMCD sum rules \cite{Thole.B_etal.Phys.-Rev.-Lett.1992,Chen.C_etal.Phys.-Rev.-Lett.1995}.  The $3d$ electron occupation number appearing in the sum rule was set to 5. The estimated spin and orbital magnetic moments per Mn ion are  $m_{\text{spin}}$ = 0.45 $\mu_{\text{B}}$ and $\mu_{\text{orb}}$ = 0.05 $\mu_{\text{B}}$, respectively.  We attribute the small spin moment compared with 5 $\mu_{\text{B}}$, the value expected for Mn$^{2+}$ ($d^{5}$), to a competition between the antiferromagnetic (AFM) superexchange interaction for the nearest-neighbor Mn atoms and ferromagnetic (FM) interaction through long-range coupling mediated by the itinerant carriers. A theoretical work predicts that the effective exchange coupling $J_{H}^{\text{eff}}$ between Mn pairs changes sign from AFM to FM when they are separated by more than 3 \AA\ for hole-doped cases, $x=0.2$ and 0.4 \cite{Glasbrenner.J_etal.Phys.-Rev.-B2014}. Indeed, the end compound BaMn$_{2}$As$_{2}$ is a local-moment AFM insulator with the extraordinarily high N\'eel temperature of $T_{N}=$ 625 K, and the AFM order remains robust in the hole-doped system Ba$_{1-x}$K$_{x}$Mn$_{2}$As$_{2}$ up to $x=0.40$ \cite{Pandey.A_etal.Phys.-Rev.-Lett.2012,Yu.T_etal.2020}. The XMCD magnetic moments are consistent with the total magnetization estimated from SQUID magnetometry \cite{Zhao.G_etal.Sci.-Rep.2017}. %A possible cause for the discrepancy is the ferromagnetic polarization of itinerant As $4p$ holes which mediate the magnetic interaction between the Mn $3d$ local spins.
The small orbital magnetic moment is in line with GaMnAs \cite{Takeda.Y_etal.Phys.-Rev.-Lett.2008}, reaffirming the $S=5/2$ high spin configuration of Mn $3d$ electrons. The nonzero value reflects the finite contribution from the charge-transferred states in the ground state.

\begin{figure}[htbp] %  figure placement: here, top, bottom, or page
   \centering
   \includegraphics[width=8cm]{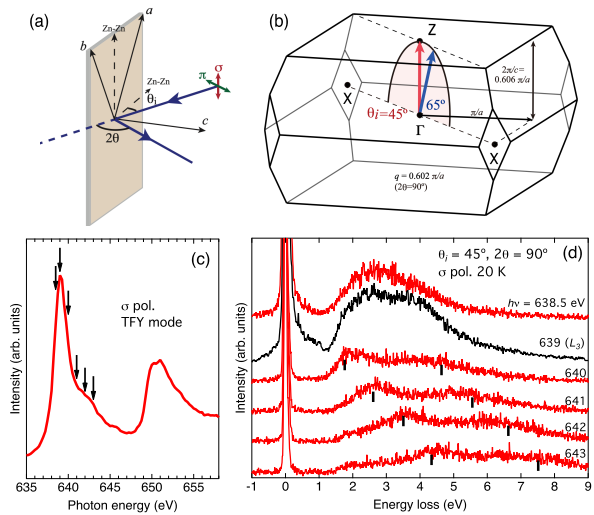} 
%   \caption{(Color online) (a) Valence-band ARPES difference spectrum of Mn-BaZn$_{2}$As$_{2}$. The off-resonance ARPES spectrum taken at 637 eV is subtracted from the on-resonance one at 640 eV in order to highlight the resonance enhancement of spectral weight. No significant enhancement due to the Mn $3d$ impurity band was observed near $E_{F}$. (b) Mn $3d$ partial density of states obtained by integrating panel (a) over momenta. Off- and on-resonance photoemission spectra obtained from Figs. \ref{RPES} (b) and (c) are also shown.}
   \caption{(Color online) (a) Scattering geometry of the resonant inelastic x-ray scattering (RIXS) experiment. (b) Momentum transfer ${\bm q}$ in the Brillouin zone of Ba$_{1-x}$K$_{x}$(Zn$_{1-y}$Mn$_{y}$)$_{2}$As$_{2}$. (c) XAS spectra taken in the total-fluorescence-yield mode. The incident photon energies employed for RIXS measurements are indicated by arrows. (d) Incident energy dependence of the RIXS spectra taken with the $\sigma$ polarization. Vertical bars indicate fluorescent-like emission that has a constant x-ray energy.}
   \label{RIXS}
\end{figure}

%%RIXS%%
Having established the magnetic properties of the Mn ions in the ground state, we study the electronic configurations via the excitation spectra. Figure \ref{RIXS} (a) shows the scattering geometry of the RIXS experiment. The detector angle $2\theta$ was fixed at 90$^{\circ}$ throughout this work, and the polarization of incident photons was kept either within or perpendicular to the scattering plane (referred to as $\pi$ and $\sigma$ scattering geometries, respectively).  The momentum transfer ${\bm q}$ in the Brillouin zone of Ba$_{1-x}$K$_{x}$(Zn$_{1-y}$Mn$_{y}$)$_{2}$As$_{2}$ is shown in panel (b). At the Mn $L_{3}$ edge with $2\theta=90^{\circ}$ geometry, the magnitude of ${\bm q}$ was 0.602 $\pi/a$ or 0.993 $2\pi/c$, where $a=4.12$ \AA\ and $c=13.58$ \AA\ are in-plane and out-of-plane lattice constants, respectively. The incident angle $\theta_{i}$ with respect to the $a$-$b$ plane can be varied by rotating the sample on the rotation axis perpendicular to the scattering plane. In the momentum space, this corresponds to rotating the ${\bm q}$ vector within the $\Gamma$-XZ plane as illustrated in the figure. In the specular scattering geometry ($\theta_{i}=45^{\circ}$), the ${\bm q}$ vector approximately corresponds to the $\Gamma$-Z distance ($2$$\pi/c=0.606$ $\pi/a$).

Figure \ref{RIXS} (c) shows XAS spectra in the Mn $L_{2,3}$-edge region of Mn-BaZn$_{2}$As$_{2}$ collected in the total-fluorescence-yield mode. The Mn $L_{3}$ peak is located at 639 eV and the photon energies used in the following RIXS measurements are indicated by black arrows. Figure \ref{RIXS} (d) shows RIXS spectra taken with various incident photon energies ($h\nu$) across the $L_{3}$ peak with the $\sigma$ polarization. %Due to the diluteness of the Mn concentration, the count rate for each energy point around -3 eV energy loss was $\sim$ 30 counts per hour at the $L_{3}$ peak at -3 eV and the data were accumulated for about 6 hours for each spectrum. 
At the $L_{3}$ peak, we observed a broad feature between 1 eV and 6 eV energy loss. This broad feature originates from the $d$-$d$ excitation of the Mn ion. Upon detuning the photon energy from the $L_{3}$ peak toward higher energies, we observe broad continuous excitations that exhibit fluorescent behavior. These excitations originate from emission from the intermediate states where electrons in the As $4p$ valence band screen the Mn $2p$ core holes through the As $4p$-Mn $3d$ hybridization.

\begin{figure}[htbp] %  figure placement: here, top, bottom, or page
   \centering
   \includegraphics[width=8cm]{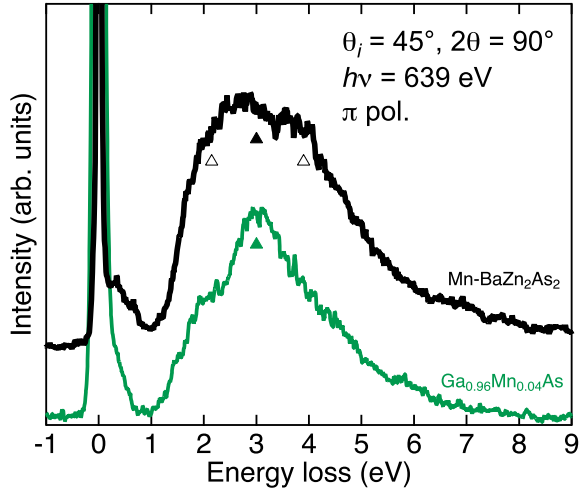} 
%   \caption{(Color online) (a) Valence-band ARPES difference spectrum of Mn-BaZn$_{2}$As$_{2}$. The off-resonance ARPES spectrum taken at 637 eV is subtracted from the on-resonance one at 640 eV in order to highlight the resonance enhancement of spectral weight. No significant enhancement due to the Mn $3d$ impurity band was observed near $E_{F}$. (b) Mn $3d$ partial density of states obtained by integrating panel (a) over momenta. Off- and on-resonance photoemission spectra obtained from Figs. \ref{RPES} (b) and (c) are also shown.}
   \caption{(Color online) Comparison of RIXS spectra for Ba$_{0.7}$K$_{0.3}$(Zn$_{0.85}$Mn$_{0.15}$)$_{2}$As$_{2}$  and Ga$_{0.96}$Mn$_{0.04}$As \cite{Kobayashi.M_etal.Phys.-Rev.-Lett.2014}. The spectra are taken with $\pi$-polarized photons at the Mn $L_3$ peak. Filled triangles indicate the peak originating from the Mn$^{3+}$ configuration, and open triangles indicate those from the Mn$^{2+}$ configuration.}
   \label{RIXS_CI}
\end{figure}

In Fig. \ref{RIXS_CI}, we compare the RIXS spectrum taken at the Mn $L_{3}$ peak with the $\pi$ polarization and that for Ga$_{0.96}$Mn$_{0.04}$As from Kobayashi \textit{et al}. \cite{Kobayashi.M_etal.Phys.-Rev.-Lett.2014}. In both systems, the $d$-$d$ excitation profiles are much broader than the experimental energy resolution of 80 meV. As an origin of the broadening, we consider the fast decay of the final states into electron-hole pairs in the As 4$p$ valence and
conduction bands, as theoretically proposed for Mn impurity in Ag \cite{Taguchi.M_etal.Phys.-Rev.-B2006}. In Ga$_{0.96}$Mn$_{0.04}$As, the small band gap of the host GaAs opens a decay channel for the $d$-$d$ excitations into the continuum of electron-hole pairs in the GaAs host \cite{Kobayashi.M_etal.Phys.-Rev.-Lett.2014}. The same decay mechanism is expected in Mn-BaZn$_{2}$As$_{2}$, as  the narrower band gap of 0.23 eV of BaZn$_{2}$As$_{2}$ \cite{Xiao.Z_etal.Journal-of-the-American-Chemical-Society2014} than that of GaAs (1.42 eV) makes the electron-hole pair creation more efficient. Regarding the peak positions of the $d$-$d$ excitations, the main peak of Ga$_{0.96}$Mn$_{0.04}$As is located at 3 eV (green triangle), which corresponds to the peak in the calculated spectra for the Mn$^{3+}$ state. Kobayashi \textit{et al}. concluded that the charge-transferred $3d^{5}\underline{L}$ state is more appropriate for the description of the Mn impurity state in GaMnAs. For Mn-BaZn$_{2}$As$_{2}$, on the other hand, peak structures are also present around 2 and 4 eV (open triangles), which correspond to the peaks in the calculated spectra for the Mn$^{2+}$ state. In addition, the intensity between 2 and 4 eV (black triangle) is also high, which is attributed to emission from the Mn$^{3+}$ configuration as in GaMnAs. Based on this consideration, we conclude that both the pure acceptor $3d^{5}$ electron configuration with free holes and the $3d^{5}\underline{L}$ electron configuration with bound holes contribute to the ground state of Mn ions in Mn-BaZn$_{2}$As$_{2}$.

\begin{figure}[htbp] %  figure placement: here, top, bottom, or page
   \centering
   \includegraphics[width=8cm]{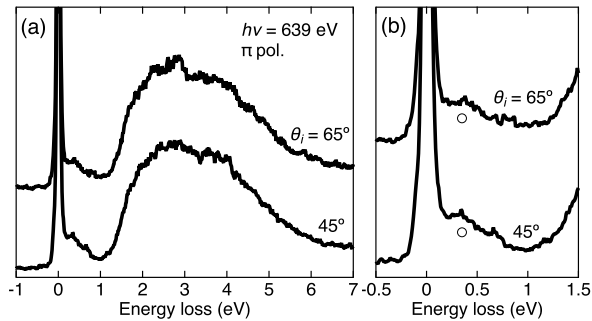} 
%   \caption{(Color online) (a) Valence-band ARPES difference spectrum of Mn-BaZn$_{2}$As$_{2}$. The off-resonance ARPES spectrum taken at 637 eV is subtracted from the on-resonance one at 640 eV in order to highlight the resonance enhancement of spectral weight. No significant enhancement due to the Mn $3d$ impurity band was observed near $E_{F}$. (b) Mn $3d$ partial density of states obtained by integrating panel (a) over momenta. Off- and on-resonance photoemission spectra obtained from Figs. \ref{RPES} (b) and (c) are also shown.}
   \caption{(Color online) (a) RIXS spectra taken at the Mn $L_{3}$ edge with $\pi$ polarizations for incident angles $\theta_{i}=45^{\circ}$ and $65^{\circ}$. (b) Enlarged plot of the low energy-loss region. Circles indicate broad peaks originating from the Stoner excitations.}
   \label{qdep}
\end{figure}

In addition to the $d$-$d$ excitation of relatively high energies primarily localized on the Mn ion, it is important to study low-energy magnetic excitations. In order to investigate such excitations, the excitations in the lower energy region were measured with $\pi$ polarization for incident angles $\theta_{i}=45^{\circ}$ and $65^{\circ}$, as shown in Fig. \ref{qdep} (a).  In the scattering geometry with $2\theta=90^{\circ}$, the $\pi$ polarization vector of the incident photons is always orthogonal to the two polarization vectors of the scattered photons. This condition suppresses elastic charge scattering \cite{Blume.M_etal.J.-Appl.-Phys.1985} and prevents the high-energy tail of the elastic peak from obscuring low-energy excitations.  Instead, the RIXS response is dominated by the magnetic excitations. The broad $d$-$d$ peaks do not show significant change between $\theta_{i}=45^{\circ}$ and $65^{\circ}$, as expected for the local excitations.  Figure \ref{qdep} (b) shows an expanded plot of the low-energy region. In addition to the elastic peak and the $d$-$d$ excitations, we observe weak and broad peaks around 0.35 eV as indicated by open circles. Theoretical studies based on the $p$-$d$ exchange model predict that ferromagnetic DMSs can support collective acoustic and optical spin-wave modes, and a continuum of Stoner excitations \cite{Konig.J_etal.Phys.-Rev.-B2001,Konig.J_etal.Phys.-Rev.-Lett.2000,Konig.J_etal.Physica-E2001}. Their energy scales are determined by the zero-temperature spin-splitting gap $\Delta=cJ_{pd}S$ for the itinerant carriers, where $c$ is the magnetic ion density, $J_{pd}$ is the $p$-$d$ exchange and $S=5/2$ is the local spin. As typical parameters for GaMnAs, K\"onig \textit{et al.} \cite{Konig.J_etal.Phys.-Rev.-Lett.2000} take $c=1$ nm$^{-3}$ and $J_{pd}=$ 0.15 eV nm$^{3}$. Then the energy scale of the acoustic spin wave becomes $0.02 \Delta\sim$10 meV, which is difficult to detect with the present energy resolution. On the other hand, the optical spin waves and Stoner excitations occur at much higher energy scales $\Delta$ ($\sim$ 0.4 eV) because the energy scales are determined by the flipping of a single spin in the ferromagnetic itinerant carriers. This energy scale matches that of the weak peak at 0.35 eV. The direct observation of the acoustic spin wave requires further RIXS measurements of the low energy region with higher resolution.

%%Cluster Model%%

%%%ADD Discussion on composition between Mn2+ and Mn3+ !!!

%%Discussion%%

%The saturated magnetic moment per Mn observed here is significantly reduced from 5 $\\mu_{\text{B}$ expected from the high-spin configuration of Mn $3d$ electrons. Reference \onlinecite{Glasbrenner.J_etal.Phys.-Rev.-B2014} ascribes this to the competition between the AFM superexchange coupling between the nearest-neighbor Mn spins and the longer-range FM coupling between Mn spins mediated by As holes. The fact that the end material BaMn$_{2}$As$_{2}$ is an AFM insulator with a high N\'eel temperature ($T_{N}$) of 625 K and that Mn remains AFM in the hole-doped metallic system Ba$_{1-x}$K$_{x}$Mn$_{2}$As$_{2}$ with a slight reduction of $T_{N}$ (480 K for $x$ $\sim$ 0.4) \cite{Pandey.A_etal.Phys.-Rev.-Lett.2012} is in accordance with this picture. It has been found \cite{Pandey.A_etal.Phys.-Rev.-Lett.2013,Ueland.B_etal.Phys.-Rev.-Lett.2015} that the As $4p$ conduction band in Ba$_{1-x}$K$_{x}$Mn$_{2}$As$_{2}$ shows itinerant ferromagnetism and coexists with the Mn AFM order. 

%%Conclusion%%
In conclusion, we have investigated the magnetic properties and the electronic configurations of Mn ions in the layered ferromagnetic semiconductor Mn-BaZn$_{2}$As$_{2}$ using the XMCD and RIXS techniques. Ferromagnetic XMCD signal from the Mn $L_{3}$ edge is observed with spin moment of 0.45 $\mu_{\text{B}}$/Mn and small orbital moment of 0.05 $\mu_{\text{B}}$/Mn, indicating that the Mn atoms take the high-spin $3d^{5}$ electron configuration and that there are competing FM and AFM interactions between the local moments. Mn $L_{3}$ edge RIXS spectra show a broad peak between 1 to 6 eV energy loss, which is assigned to the $d$-$d$ excitations from the Mn ground states.  The broadness of this feature originates from the rapid decay of the $d$-$d$ excitons to the electron-hole pairs in the As $4p$ valence and conduction bands with a small indirect band gap.
 The RIXS line shape indicates that the ground state of Mn in Mn-BaZn$_{2}$As$_{2}$ consists not only of the charge-transferred $3d^{5}\underline{L}$ electron configuration with weakly bound itinerant holes, but also of the pure $3d^{5}$ configuration with free holes.

This work was supported by Grants-in-Aid for Scientific Research from JSPS (S22224005, 19K03741), Spintronics Research Network of Japan (Spin-RNJ), Reimei Project from Japan Atomic Energy Agency, Nanotechnology Platform (project No.12024606) 
of the MEXT,
Friends of Todai Foundation in New York, the US NSF PIRE (Partnership for International Research and Education: OISE-0968226), and the US National Science Foundation Grant No. DMR-1610633. XMCD and RIXS experiments at NSRRC were performed under the proposals No. 2012-2-086-4 and 2015-3-115-1. Work at IOPCAS was supported by NSF \& MOST as well as CAS International Cooperation Program of China through Research Projects. G.Q.Z. has been supported in part by China Scholarship Council (No. 201904910900). H.S. acknowledges financial support from Advanced Leading Graduate Course for Photon Science (ALPS) and the JSPS Research Fellowship for Young Scientists. 

\bibliography{XMCD_RIXS_DMS}

\end{document}